%% file: main.tex
\documentclass{sig-alternate-05-2015}

\usepackage[bookmarks=false, hidelinks]{hyperref}
\usepackage{subcaption}
\usepackage[vlined, ruled]{algorithm2e}
\SetProcNameSty{textsc}

\renewcommand*{\algorithmautorefname}{Algorithm} % force capitalization

\usepackage{paralist}

\usepackage{multirow}

\usepackage{tikzexternal}
\usepackage{pgfplots}
\usepgfplotslibrary{external}
\usepackage{url}

\pgfplotsset{
	jitter/.style={
		x filter/.code={\pgfmathparse{\pgfmathresult+rnd*#1}}
	},
	jitter/.default=0.1
}

\begin{document}

% Copyright
%\setcopyright{acmcopyright}
%\setcopyright{acmlicensed}
\setcopyright{rightsretained}
%\setcopyright{usgov}
%\setcopyright{usgovmixed}
%\setcopyright{cagov}
%\setcopyright{cagovmixed}

% DOI
\doi{}

% ISBN
\isbn{}

%Conference
\conferenceinfo{}{}

\acmPrice{\$0.00}

%
% --- Author Metadata here ---
\conferenceinfo{}{}
%\CopyrightYear{2007} % Allows default copyright year (20XX) to be over-ridden - IF NEED BE.
%\crdata{0-12345-67-8/90/01}  % Allows default copyright data (0-89791-88-6/97/05) to be over-ridden - IF NEED BE.
% --- End of Author Metadata ---

\title{Finer-grained Locking in Concurrent \\Dynamic Planar Convex Hulls}

\numberofauthors{2} %  in this sample file, there are a *total*
\author{
\alignauthor
K. Alex Mills\\
\affaddr{UT Dallas -- ANDES Lab}\\
\affaddr{800 W. Campbell Rd.}\\
\affaddr{Richardson, TX USA}\\
\email{k.alex.mills@utdallas.edu}
\alignauthor
James Smith\\
	   \affaddr{UT Dallas -- ANDES Lab}\\
       \affaddr{800 W. Campbell Rd.}\\
       \affaddr{Richardson, TX USA}\\
       \email{james@nullious.net}
}
% There's nothing stopping you putting the seventh, eighth, etc.
% author on the opening page (as the 'third row') but we ask,
% for aesthetic reasons that you place these 'additional authors'
% in the \additional authors block, viz.
\date{9 Feb 2017}
% Just remember to make sure that the TOTAL number of authors
% is the number that will appear on the first page PLUS the
% number that will appear in the \additionalauthors section.

\maketitle
\begin{abstract}
The convex hull of a planar point set is the smallest convex polygon 
containing each point in the set. The dynamic convex hull problem
concerns efficiently maintaining the convex hull of a set of points subject to additions and removals.
One algorithm for this problem uses two external balanced binary search trees (BSTs) \cite{Overmars1981}. We present the first concurrent solution for this problem, which uses a \emph{single} BST that stores references to intermediate convex hull solutions at each node. We implement and evaluate two lock-based approaches: \begin{inparaenum}[a)] \item fine-grained locking, where each node of the tree is protected by a lock, and \item ``\emph{finer}-grained locking," where each node contains a separate lock for each of the left and right chains. \end{inparaenum}% 
In our throughput experiments, we observe that finer-grained locking yields an 8-60\% improvement over fine-grained locking, and a 38-61$\times$ improvement over coarse-grained locking and software transactional memory (STM). When applied to find the convex hull of \emph{static} point sets, our approach outperforms a parallel divide-and-conquer implementation by 2-4$\times$ using an equivalent number of threads.
\end{abstract}




%
%  Use this command to print the description
%
%\printccsdesc

% We no longer use \terms command


\section{Introduction}
Historically, research on concurrent data structures has focused on general-purpose solutions which find common application (e.g. linked lists, skip-lists, binary search trees). We break from this tradition to focus on a specific problem: maintaining the convex hull of a \emph{dynamic} planar point-set subject to insertions and deletions.

In the sequential setting, the planar convex hull problem is well-studied. There exist well-known algorithms which match the lower-bound of $\Omega(n \log n)$ \cite{Graham1972, Andrew1979} and output-sensitive algorithms which run in $O(n \log h)$ time \cite{Kirkpatrick1986, Chan1996}, where $h$ is the number of points on the hull.

In the dynamic setting \cite{Preparata1979}, Overmars and van Leeuwen presented an algorithm in which inserts and deletes can be achieved in $O(\log^2 n)$ time \cite{Overmars1981}. Later algorithms improved on this foundation, \cite{Chan2001, Brodal2002} achieving $O(\log n )$ time for both inserts and deletes \cite{Brodal2002}, but they are primarily of theoretical interest. The dynamic variant has various applications \cite{Bespamyatnikh2000, Tamir1988, Chazelle1987}, including k-nearest neighbor search \cite{Edelsbrunner1986}, least-squares clustering \cite{Aurenhammer1998}, and sketching geometric points \cite{Hershberger08}.

\section{Dynamic Convex Hulls}\label{s:convex-hull}

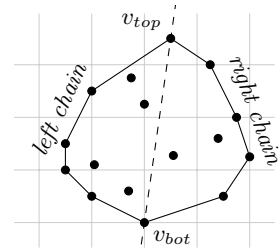
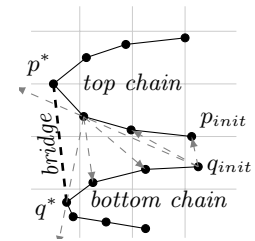
\begin{figure}[b]
	\centering
	\begin{subfigure}{0.45\linewidth}
		\centering
		\input{fig-convex-poly-hull}
		\subcaption{A convex hull}\label{f:convex-poly-hull}
	\end{subfigure}
	\hfill
	\begin{subfigure}{0.45\linewidth}	
		\centering
		\input{fig-bridge} 
		\subcaption{Bridge-finding example (only left chains are shown).}\label{f:bridge}
	\end{subfigure}
	\caption{Terminology and bridge-finding example.}
\end{figure}

Throughout computational geometry it is standard practice to require that point sets are specified in \emph{general position} w.l.o.g. \cite{deBerg2010}. In this paper and our experiments, general position means that \begin{inparaenum}[a)] \item no three points are co-linear, and \item no two points share a horizontal line. \end{inparaenum} 

A \emph{convex polygon} is a simple polygon in which every internal angle is acute. The \emph{convex hull} of a planar point set $S \subset \mathbb{R}^2$, denoted $conv(S)$, is the smallest convex polygon which contains all of the points in $S$. We use variables $n$ and $h$ to denote $n = |S|$ and $h = |conv(S)|$. In memory, we store convex hulls as a pair of linked lists, called the right and left chains. Each chain is a linked list of vertices $v_1,v_2,...,v_h$ stored in \emph{clockwise (c.w.) order}. Vertices adjacent in the list are joined by an edge; the edge from $v_h$ to $v_1$ is implicitly included. If $v_{top}$ and $v_{bot}$ are the top and bottom-most points of $S$, then the \emph{right chain} is the list of vertices in $conv(S)$ that lie on or to the right of the line from $v_{top}$ to $v_{bot}$; the \emph{left chain} is the list of vertices which lie on or to the left of this line (\autoref{f:convex-poly-hull}).

Formally, the \emph{dynamic (planar) convex hull problem} is the specification of an abstract data type which maintains a \mbox{finite} set of points $S \subset \mathbb{R}^2$ and provides the following three operations,
\begin{inparaenum}[1)]
	\item \textsc{insert}$(p)$: add point $p \in \mathbb{R}^2$ to $S$,
	\item \textsc{delete}$(p)$: remove point $p \in \mathbb{R}^2$ from $S$ (if present),
	\item \textsc{getHull}$()$: return the convex hull of $S$.
\end{inparaenum}

\paragraph{Divide-and-conquer for Convex Hulls}\label{s:divide-and-conquer}

We modify the approach of \cite{Overmars1981}, which is based on the following divide-and-conquer algorithm. Given a \emph{static} set of points $S \subset \mathbb{R}^2$, we compute the left and right chains of the convex hull separately. For concision, we describe only the left chain w.l.o.g.; the right chain is symmetric. 

The divide step recursively splits $S$ into two equal halves, $S_{top}$ and $S_{bot}$, by the horizontal line formed by the median $y$-coordinate of $S$. The recursion terminates when two or fewer points $\{p_1,p_2\}$ remain. The \emph{left} chain of $\{p_1, p_2\}$ is the points sorted in \emph{increasing} order of $y$-coordinate (sorting ensures the c.w. ordering of the left chain). The conquer step combines the left chains of $S_{top}$ and $S_{bot}$ to form the left chain of their union \mbox{$S_{top} \cup S_{bot}$}. To combine the left chains of $S_{top}$ and $S_{bot}$, we find and add the \emph{bridge}, the unique line segment tangent to both chains. In \autoref{f:bridge}, the bridge is the segment between $p^\ast$ and $q^\ast$. The vertices $p^\ast$ and $q^\ast$ define the bridge and can be found by walking the top chain in clockwise order and walking the bottom chain in counter-clockwise order (\autoref{a:bridge-finding}).

\begin{algorithm}[h]
	\footnotesize
	\KwIn{a top and bottom chain separated by horiz. line}
	$p \gets$ bottom-most point of top chain ($p_{init}$)\;
	$q \gets$ top-most point of bottom chain ($q_{init}$)\;
	\Repeat{$(p,q)$ tangent to top and bottom chain} {
		\While{$(p,q)$ not tangent to top chain}{
			$p \gets$ next point on top chain in c.w. order\;
		}
		\While{$(p,q)$ not tangent to bottom chain}{
			$q \gets$ next point on bottom chain in c.c.w. order\;
		}
	}
	\Return{$(p,q)$}\;
	
	\caption{Bridge-finding algorithm}\label{a:bridge-finding}
\end{algorithm}

The dashed lines in \autoref{f:bridge} depict the tangents tested in the while loops of \autoref{a:bridge-finding}. Once the bridge is found, the lists representing the top and bottom chains are \emph{split} at the points $p^\ast$ and $q^\ast$. The (left) chain of $S_{top} \cup S_{bot}$ is formed by contatenating the part of the top chain above and including $p^\ast$ with the bottom chain below and including $q^\ast$. 

The parallel convex hull strategy we test against uses this algorithm. To avoid expensive median-finding, we sort by $y$-coordinate using Arrays.parallelSort() \cite{Oracle2014Arrays}. Each recursive call on a subproblem with more than 2000 points is submitted to a ForkJoinPool \cite{Oracle2014}. Recursive calls on fewer than 2000 points are solved sequentially. Cutting off parallelism after 2000 points yields good performance for our experimental workload in particular.

\paragraph{Dynamic Convex Hulls using BSTs}

Following \cite{Overmars1981}, we store the solutions to divide-and-conquer subproblems at the internal nodes of a \emph{HullTree}, an external BST in which each (internal) node has either zero or two children. Leaf nodes store the points of $S$ sorted in order by y-coordinate. We refer to the two children of an internal node as the \emph{top} and \emph{bottom} children, which store both chains of the top and bottom subproblems respectively. Each node $u$ stores a reference to the left and right chains of all the points stored at the leaf nodes below $u$. Just as in BSTs, we store the minimum key from the top subtree to aid in routing search operations.

Inserting or deleting point $p$ from the HullTree occurs as usual except it is immediately followed by a leaf-to-root pass back up the tree to recompute the chains. During this pass, if $u$ has children $c_1$ and $c_2$, we execute the conquer step on both chains stored at $c_1$ and $c_2$, and \emph{replace} the chains stored at node $u$ with \emph{copies of} the new chains which result. The leaf-to-root pass continues at $u$'s parent. This procedure propagates all needed changes to the root node; it recomputes only the subproblems whose solutions could change.

\begin{comment}
\begin{figure}
	\centering
	\input{fig-ovl-naive-illustration}
	\caption{HullTree illustration (only the left chain is shown). Bolded edges depict the search path for point $c$.}\label{f:ovl-naive-illustration}
	\vspace{-0.15cm}
\end{figure}
\end{comment}

\section{Finer-grained Locking}

To manage concurrency in the ``\emph{finer}-grained locking" strategy, we store a \emph{leftLock}, a \emph{rightLock}, and an \emph{isDeleted} flag at each node in addition to the child, parent, and chain references mentioned in \autoref{s:divide-and-conquer}. $leftLock$ and $rightLock$ are used to protect the left and right chains, respectively. We also use $leftLock$ to protect access to a node's parent and child pointers. The fine-grained variant uses only one lock to protect access to all of a node's data members.

\begin{figure*}[t]
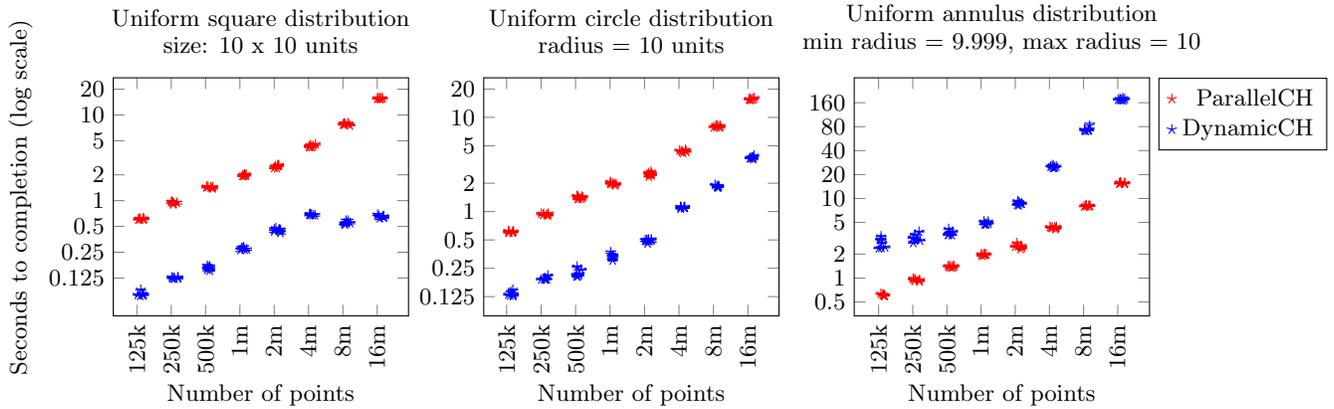
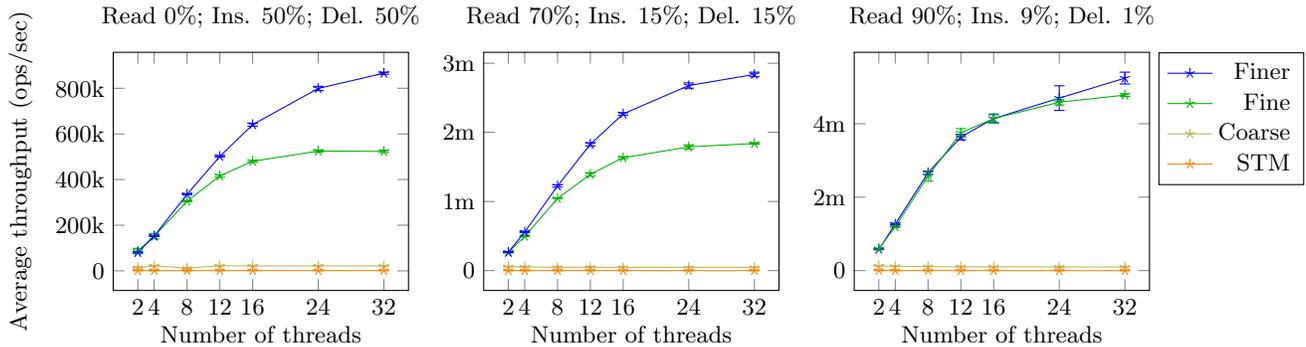

	\centering
	\begin{subfigure}{\linewidth}
		\include{parallel-groupplot}
		\vspace{-12pt}
		\subcaption{Six instances of each size were averaged over 10 runs. Random noise added to $x$-coordinates for visual clarity.}\label{f:static-exps}
		\vspace{5pt}
	\end{subfigure}
	\begin{subfigure}{\linewidth}
		\include{throughput-groupplot}
		\vspace{-12pt}
		\subcaption{Each point averages six, 15-second runs. Measurements taken after a 2-second warmup period. 95\% confidence intervals are shown.}\label{f:throughput-exps}
		\vspace{5pt}
	\end{subfigure}
	\vspace{-12pt}
	\caption{Experimental results.}
	\vspace{-12pt}
\end{figure*}

\SetKwFunction{merge}{merge} 

In both strategies, reads only need to access the convex hull stored at the root. We describe the procedure for writes as follows. Writes involving point $p$ begin with a search down the HullTree to find the leaf node closest to $p$. During this search, locks are not acquired. Let $u$ be the leaf node found by a search for point $p$.

In case of an insert, we lock $u.leftLock$. Next, we validate to ensure \begin{inparaenum}[a)]\item that $u$ is still a leaf node, and \item that $u.isDeleted$ is false\end{inparaenum}. If validation fails, the insert operation restarts from the root. Otherwise, the leaf receives two newly allocated child nodes, one which contains $p$ and another which contains the point stored at $u$. Then, both chains of $u$ are updated so that they are the left/right chains of the points contained at $u$'s children. Finally, \mbox{$u.leftLock$} is released and \merge{$u$} is invoked (\autoref{a:merge}).

In case of a delete, we acquire a three-node window of locks in this sequence: \begin{inparaenum}[1)]
	\item we lock $u.leftLock$,
	\item we \emph{attempt} to lock the $leftLock$ of $u$'s sibling, denoted $u_{sib}$,
	\item we lock the $leftLock$ of $u$'s parent, denoted $u_{par}$.
\end{inparaenum}
Thus locks are acquired bottom-up. If the attempt to lock $u_{sib}$ fails, locks are released, and the operation is retried from the root. Once all locks are acquired, the window is validated to ensure 
\begin{inparaenum}[1)]
	\item that none of $u$, $u_{sib}$, or $u_{par}$ have their $isDeleted$ flag set to true, and
	\item that $u_{sib}$ and $u$ are still children of $u_{par}$.
\end{inparaenum} 
If validation fails, the operation is retried from the root. Finally, the delete target is effectively removed from the tree via the following sequence of updates 
\begin{inparaenum}[1)] 
	\item we replace $u_{par}$'s left/right chains by the left/right chain of $u_{sib}$, 
	\item we redirect $u_{par}$'s top/bottom child pointers to point to the top/bottom children of $u_{sib}$, denoted $c_1$ and $c_2$, 
	\item we redirect the parent pointers of $c_1$ and $c_2$ to refer to $u_{par}$, and 
	\item we set $u$ and $u_{sib}$'s $isDeleted$ flag to true.
\end{inparaenum}
Finally, all locks are released and \merge{$u_{par}$} is invoked (see \autoref{a:merge}).

The merge operation (\autoref{a:merge}) performs ``finer-grained hand-over-hand locking" up the tree, allowing updates to the left and right chains to occur concurrently while ensuring no pair of merge operations cross on their way to the root. Since $leftLock$ is acquired first, if a merge collides with a write, the merge cannot proceed until the write has released its locks. Since writes only acquire leftLocks, it is possible for a write operation to overlap a merge while the merge is updating the right chain. Since writes are immediately followed by a merge, this does not yield inconsistencies. We implement a minor optimization in which merges stop early once the chains stop changing as a result of the updates.

\begin{algorithm}[h]
	
	\footnotesize
	\SetKwProg{proc}{Procedure}{}{}
	
	\proc{\merge{$node$}}{
		$prev \gets null$\;
		\While{$node \neq null$} {
			$node.leftLock.lock()$\;
			\lIf{$prev \neq null$}{$prev.rightLock.unlock()$}
			$\{~... \text{ update left chain } ...~\}$ \DontPrintSemicolon \; \PrintSemicolon
			$node.rightLock.lock()$\;
			$node.leftLock.unlock()$\;	
			$\{~... \text{ update right chain } ...~\}$ \DontPrintSemicolon \; \PrintSemicolon
			$prev \gets node$; $node \gets node.parent$\;
		}
		\lIf{$prev \neq null$}{$prev.rightLock.unlock()$}	    
	}
	\caption{Merge operation.}\label{a:merge}
\end{algorithm}
	\vspace{-0.5cm}

\paragraph{Safety and Consistency Properties}

Our approach is easily seen to be deadlock-free, since 
\begin{inparaenum}[1)]
	\item locks are always acquired bottom-up, and 
	\item $leftLock$ is always acquired before $rightLock$.
\end{inparaenum}
In finer-grained locking, reads may find the root node in an inconsistent state when the top/bottommost points of the hull change due to concurrent writes. This condition may be tested for by ensuring that the top/bottommost points in the left and right chains match. If they do not, the situation can be remedied by 
\begin{inparaenum}[(a)]
	\item retrying until a clean read occurs (linearizability),
	\item taking the convex hull of the result in $O(h)$ time (quiescent consistency), or
	\item returning a possibly inconsistent view to the caller.
\end{inparaenum}
Since inconsistent intermediate views are acceptable in \emph{static} workloads, we use approach (c).

\section{Experiments and Conclusions}

\autoref{f:static-exps} depicts three experiments on static instances, each run using 24 threads. They were performed on a 64-bit 12-core AMD Opteron 6180 SE with 64 GB of RAM, (24 hyperthreads). ``ParallelCH" is the parallel divide-and-conquer implementation previously mentioned. ``DynamicCH" tests each point to see if it lies in the convex hull of the finer-grained concurrent data structure. If the point is not already in the convex hull, it is inserted into the data structure, otherwise it is ignored. All implementations are in Java. In the left and middle plots, instances are formed by drawing points uniformly at random from a box and a circle respectively. The right plot is a stress test in which points are sampled from an extremely thin annulus (area: 0.062 units$^2$). This scenario is \emph{\textbf{far from a typical case}}: the sampled points are extremely close to the boundary in a \emph{tiny} region measuring only $1/5000 \times$ the area of the circle used in the middle plot. This distribution forces the dynamic data structure to manage many large chains. Yet even in this case, using finer-grained locking yields a throughput speedup of between 8-60\%. In \autoref{f:throughput-exps} three workloads \emph{using the same annulus distribution} are shown. These experiments were performed on a machine using two 64-bit 16-core Intel Xeon E5-4650 processors with 64GB of RAM (32 threads). DeuceSTM v1.3 was used in the STM variant \cite{Korland2010}.

In future work, we propose investigating balanced HullTrees and the space-saving techniques described in \cite{Overmars1981}.

\section{Acknowledgements}

We are grateful to Neeraj Mittal, Shreyas Gokhale, and Kenneth Platz for insightful conversations and feedback on the design and experimental analysis phases of this project. We would also like to thank Shreyas Gokhale, Joel Long, and Gregory van Buskirk for their constructive feedback on drafts.

\bibliographystyle{abbrv}
\bibliography{spaa2017}  % sigproc.bib is the name of the Bibliography in this case
% ACM needs 'a single self-contained file'!
%\balancecolumns % GM June 2007
% That's all folks!
\end{document}

%% file: fig-convex-poly-hull.tex
\begin{tikzpicture}[scale=0.35,
 extended line/.style={shorten >=-#1,shorten <=-#1},
extended line/.default=1cm]

\draw[step=2, help lines, black!20] (-0.95,-0.95) grid (8.95,7.95);

\foreach \Point in {(2,1), (4,0),(7,1),(8,2.5),(7.5,4),(6.5,6),(5,7),(2,5),(1,3),(1,2), (4,4.5), (3.5,5.5), (2.1,2.2), (3.4,1.2), (6.8,3.2), (5.1,2.55)} 
\draw[fill = black] \Point circle (0.15) node[] {};

\draw (2,1) -- (4,0)--(7,1)--(8,2.5)--(7.5,4)--(6.5,6)--(5,7)--(2,5)--(1,3)--(1,2)--(2,1);

\draw [extended line=0.5cm, dashed] (5,7) -- (4,0);

\node at (5,7) [above left] {$v_{top}$};
\node at (4,0) [below right] {$v_{bot}$};

\node at (1.5,4) [above, rotate=65] {\emph{left chain}};
\node at (7.5,4) [above, rotate=290] {\emph{right chain}};

\end{tikzpicture}

%% file: fig-bridge.tex
\begin{tikzpicture}[scale=0.35,
extended line/.style={shorten >=-#1}]

\draw[step=2, help lines, black!20] (0.15,0.15) grid (7.95,8.95);

\foreach \Point in {(6.25,4), (3.95,4.25), (2.15,4.76), (1,6), (2.25,7), (3.75,7.5), (6,7.75)} 
\draw[fill = black] \Point circle (0.15) node[] {}; 

\draw (6.25,4)-- (3.95,4.25)-- (2.15,4.76)-- (1,6)-- (2.25,7)-- (3.75,7.5)-- (6,7.75);

\foreach \Point in {(4.5,0.5), (3, 0.75), (1.75,0.95), (1.5,1.5), (2.5,2.25), (4.5,2.75), (6.5,2.85)} 
\draw[fill = black] \Point circle (0.15) node[] {}; 

\draw (4.5,0.5)-- (3, 0.75)-- (1.75,0.95)-- (1.5,1.5)-- (2.5,2.25)-- (4.5,2.75) -- (6.5,2.85);

\node at (1.5,1.25) [left] {$q^\ast$};

\node at (1.25,6) [above left] {$p^\ast$};

\draw [dashed, line width=1pt] (1.5,1.5)  -- (1,6) node [midway, rotate=97, yshift=1.5mm] {\footnotesize \emph{bridge}};

\node at (6.25,4) [above right] {$p_{init}$};
\node at (6.5,2.85) [right] {$q_{init}$};

% tangents 
\draw [dashed, black!50, -latex] (6.5, 2.85) -- (6.25,4);
\draw [dashed, black!50,-latex] (6.5, 2.85) -- (3.95,4.25);
\draw [dashed, black!50, extended line=1cm,-latex] (6.5, 2.85) -- (2.15,4.75);

\draw [dashed, black!50,-latex] (2.15, 4.75) -- (4.5,2.75);
\draw [dashed, black!50,-latex] (2.15, 4.75) -- (2.5,2.25);
\draw [dashed, black!50, extended line=0.6cm,-latex] (2.15, 4.75) -- (1.5,1.5);

\node at (4,6) {\emph{top chain}};
\node at (5,1.65) {\emph{bottom chain}};

\end{tikzpicture}

%% file: parallel-groupplot.tex
\begin{tikzpicture}
	\begin{groupplot}[height=4.75cm, 
	 legend style={
	 	font={\footnotesize\selectfont},
		cells={anchor=east},
		legend pos=outer north east,
	},
	group style= {
		group size = 3 by 1,
		xlabels at = edge bottom,
		ylabels at = edge left},
	title style={align=center}, 
	xmode=log, 
	ymode=log, 
	xtick={125000,250000,500000,1000000,2000000,4000000, 8000000,16000000},
	xticklabels={125k, 250k, 500k, 1m, 2m, 4m, 8m, 16m},
	x tick label style={rotate=90},                        
	ytick=      {0.125, 0.25, 0.5, 1, 2, 5, 10, 20, 40, 80, 160},
	yticklabels={0.125, 0.25, 0.5, 1, 2, 5, 10, 20, 40, 80, 160},
	xlabel style={yshift=-7.5pt},
	xlabel={Number of points},
	ylabel={Seconds to completion (log scale)}]
	% BOX PLOT
	\nextgroupplot[	title={Uniform square distribution\\ size: 10 x 10 units}] 
	\addplot+[blue, only marks,jitter=0.15, mark=star, error bars/.cd, y dir=both, y explicit] 
	table [x index=1, y index=2, y error index=3, col sep=tab] 
	{AGGREGATEDFiner_box_StaticifiedFinerDenseDCH};
	\addplot+[red, only marks,jitter=0.15, mark=star, error bars/.cd, y dir=both, y explicit]
	table [x index=1, y index=2, y error index=3, col sep = tab]
	{AGGREGATED1954_box_ParallelCHList};	
	% CIRCLE PLOT
	\nextgroupplot[title = {Uniform circle distribution\\radius = 10 units}] 
	\addplot[red, only marks,jitter=0.15, mark=star, error bars/.cd, y dir=both, y explicit] 
	table [x index=0, y index=1, y error index=2, col sep=tab] 
	{AGGREGATED1954_circle_ParallelCHList};
	\addplot[blue, only marks,jitter=0.15, mark=star, error bars/.cd, y dir=both, y explicit] 
	table [x index=0, y index=1, y error index=2, col sep=tab] 
	{AGGREGATEDFiner_circle_StaticifiedFinerDenseDCH};
    % ANNULUS PLOT
	\nextgroupplot[	title={Uniform annulus distribution\\min radius = 9.999, max radius = 10}]
	\addplot+[red, only marks,jitter=0.15, mark=star, error bars/.cd, y dir=both, y explicit] 
	table [x index=0, y index=1, y error index=2, col sep=tab] 
	{AGGREGATED1954_annulus_ParallelCHList};
	\addlegendentry{ParallelCH}	
	\addplot+[blue, only marks,jitter=0.15, mark=star, error bars/.cd, y dir=both, y explicit] 
	table [x index=0, y index=1, y error index=2, col sep=tab] 
	{AGGREGATEDFiner_annulus_StaticifiedFinerDenseDCH};
	\addlegendentry{DynamicCH}
	\end{groupplot}
\end{tikzpicture}

%% file: throughput-groupplot.tex
\begin{tikzpicture}
	\begin{groupplot}[height=4.75cm, 
	 legend style={
		font={\footnotesize\selectfont},
		cells={anchor=east},
		legend pos=outer north east,
	},
group style= {
	group size = 3 by 1,
	xlabels at = edge bottom,
	ylabels at = edge left
},
	scaled ticks=false,
title style={align=center},  
xtick={2, 4, 8, 12, 16, 24, 32},
xlabel style={yshift=5pt},
xlabel={Number of threads},
ylabel={Average throughput (ops/sec)}
	]
	% R0_I50_D50
	\nextgroupplot[	title={Read 0\%; Ins. 50\%; Del. 50\% },
		            ytick={0,200000,400000,600000,800000},
		            yticklabels={0, 200k, 400k, 600k, 800k},
		 ] 
	\addplot+[green!70!black, mark=star, error bars/.cd, y dir=both, y explicit] 
	table [x index=0, y index=1, y error index=4, col sep=tab] 
%	{THROUGHPUT_R0_I50_D50_DenseDCH};
	{RESULTS-REPRO_R0_I50_D50_DenseDCH};
	\addplot+[blue, mark=star, error bars/.cd, y dir=both, y explicit] 
	table [x index=0, y index=1, y error index=4, col sep=tab] 
%	{THROUGHPUT_R0_I50_D50_FinerDenseDCH};
	{RESULTS-REPRO_R0_I50_D50_FinerDenseDCH};
	\addplot+[yellow!70!black, mark=star, error bars/.cd, y dir=both, y explicit] 
	table [x index=0, y index=1, y error index=4, col sep=tab] 
%	{THROUGHPUT_R0_I50_D50_CoarseSparseDCH};
	{RESULTS-REPRO_R0_I50_D50_CoarseSparseDCH};
	\addplot+[orange, mark=star, error bars/.cd, y dir=both, y explicit] 
	table [x index=0, y index=1, y error index=4, col sep=tab] 
%	{THROUGHPUT_R0_I50_D50_DeuceSTMSparseDCH};
	{RESULTS-REPRO_R0_I50_D50_DeuceSTMSparseDCH};
	% R70_I15_D15
	\nextgroupplot[	title={Read 70\%; Ins. 15\%; Del. 15\% },
	                ytick={0,1000000,2000000,3000000},
                    yticklabels={0, 1m, 2m, 3m}] 
	\addplot+[green!70!black, mark=star, error bars/.cd, y dir=both, y explicit] 
	table [x index=0, y index=1, y error index=4, col sep=tab] 
%	{THROUGHPUT_R70_I15_D15_DenseDCH};
	{RESULTS-REPRO_R70_I15_D15_DenseDCH};
	\addplot+[blue, mark=star, error bars/.cd, y dir=both, y explicit] 
	table [x index=0, y index=1, y error index=4, col sep=tab] 
%	{THROUGHPUT_R70_I15_D15_FinerDenseDCH};
	{RESULTS-REPRO_R70_I15_D15_FinerDenseDCH};	
	\addplot+[yellow!70!black, mark=star, error bars/.cd, y dir=both, y explicit] 
	table [x index=0, y index=1, y error index=4, col sep=tab] 
%	{THROUGHPUT_R70_I15_D15_CoarseSparseDCH};
	{RESULTS-REPRO_R70_I15_D15_CoarseSparseDCH};
	\addplot+[orange, mark=star, error bars/.cd, y dir=both, y explicit] 
	table [x index=0, y index=1, y error index=4, col sep=tab] 
%	{THROUGHPUT_R70_I15_D15_DeuceSTMSparseDCH};
	{RESULTS-REPRO_R70_I15_D15_DeuceSTMSparseDCH};
	% R90_I9_D1
	\nextgroupplot[	title={Read 90\%; Ins. 9\%; Del. 1\% },
				   	ytick={0,2000000,4000000},
                    yticklabels={0, 2m, 4m} ] 
	\addplot+[blue, mark=star, error bars/.cd, y dir=both, y explicit] 
	table [x index=0, y index=1, y error index=4, col sep=tab] 
%	{THROUGHPUT_R90_I9_D1_FinerDenseDCH};
	{RESULTS-REPRO_R90_I9_D1_FinerDenseDCH};
	\addlegendentry{Finer}	
	\addplot+[green!70!black, mark=star, error bars/.cd, y dir=both, y explicit] 
	table [x index=0, y index=1, y error index=4, col sep=tab] 
%	{THROUGHPUT_R90_I9_D1_DenseDCH};
	{RESULTS-REPRO_R90_I9_D1_DenseDCH};
	\addlegendentry{Fine}	
	\addplot+[yellow!70!black, mark=star, error bars/.cd, y dir=both, y explicit] 
	table [x index=0, y index=1, y error index=4, col sep=tab] 
%	{THROUGHPUT_R90_I9_D1_CoarseSparseDCH};
	{RESULTS-REPRO_R90_I9_D1_CoarseSparseDCH};
	\addlegendentry{Coarse}	
	\addplot+[orange, mark=star, error bars/.cd, y dir=both, y explicit] 
	table [x index=0, y index=1, y error index=4, col sep=tab] 
%	{THROUGHPUT_R90_I9_D1_DeuceSTMSparseDCH};
	{RESULTS-REPRO_R90_I9_D1_DeuceSTMSparseDCH};
	\addlegendentry{STM}	
\end{groupplot}
\end{tikzpicture}